\documentclass[10pt, twoside]{IEEEtran}

\usepackage{amsmath}
\usepackage{amssymb}
\usepackage{graphicx}
\usepackage{epsfig}
\begin{document}

\title{ RSP-Based Analysis for Sparsest
and Least $\ell_1$-Norm  Solutions to
 Underdetermined   Linear Systems  }
\author{Yun-Bin Zhao  \thanks{School of Mathematics, University of Birmingham,
Edgbaston, Birmingham B15 2TT,  United Kingdom ({\tt
y.zhao.2@bham.ac.uk}).  This work was supported by the Engineering
and Physical Sciences Research Council (EPSRC) under the grant
\#EP/K00946X/1.} \emph{IEEE member} }

 \maketitle

\begin{abstract}
  Recently, the worse-case analysis, probabilistic
analysis and empirical justification  have been employed to address
the fundamental question: When does $\ell_1$-minimization find the
sparsest solution to an underdetermined linear system? In this
paper, a  deterministic analysis, rooted in the  classic linear
programming theory, is carried out to further address this question.
We first identify a necessary  and sufficient condition for the
uniqueness of least $\ell_1$-norm solutions to linear systems. From
this condition, we deduce that a sparsest solution coincides with
the unique least $\ell_1$-norm solution to a linear system if and
only if the so-called \emph{range space property} (RSP) holds at
this solution. This yields a broad understanding of the relationship
between $\ell_0$- and $\ell_1$-minimization problems. Our analysis
indicates that the RSP truly lies at the heart of the relationship
between these two problems. Through RSP-based analysis, several
important questions in this field can be largely addressed. For
instance, how to efficiently interpret the gap between the current
theory and the actual numerical performance of $\ell_1$-minimization
by a deterministic analysis, and if a linear system has multiple
sparsest solutions, when does $\ell_1$-minimization guarantee to
find one of them? Moreover,  new matrix properties (such as the
\emph{RSP of order $K$} and the \emph{Weak-RSP of order $K$}) are
introduced in this paper, and a new theory for sparse signal
recovery based on the RSP of order $K$ is established.
\end{abstract}


% \textbf{AMS Subject Classifications:}  90C25, 90C05,  90C46, 15A06, 15A29,  94A08

\section{Introduction}
Many problems across disciplines can be formulated as the problem of
finding the sparsest solution to   underdetermined
 linear systems.   For instance,  many data types arising from signal and image
processing can be sparsely represented and the processing tasks
 (e.g., compression, reconstruction,  separation, and
transmission) often amount to
 the problem
$$ \min \{\|x\|_0: Ax=b\},$$ where $A$ is an $m\times n $ full-rank
matrix with $m<n, $  $b$ is a vector in $R^m,$  and $\|x\|_0$
denotes the number of nonzero components of the vector $x.$ The
system $Ax=b$ is underdetermined, and thus it has infinitely  many
solutions. The problem above seeking the sparsest solution to a
linear system is called $\ell_0$-minimization in the literature,  to
which even the profound linear algebra algorithms do not apply.
The $\ell_0$-minimization is known to be NP-hard \cite{N95}. From a
computational point of view,  it is natural to consider its
$\ell_1$-approximation:
$$ \min \{\|x\|_1: Ax=b\},  $$
based on which various computational methods for sparse solutions of
linear systems have been proposed (\cite{CDS98,  CWB08,
 HYZ08, YOGD08, BDE09, TW10, ZL12}). We note that the $\ell_1$ type minimization is a long-lasting
research topic in the field of numerical analysis and
 optimization (\cite{BCS78, BMR98, BFM99,
M99}). However, it has a great impact when it was first introduced
by Chen, Donoho and Saunders \cite{CDS98} in 1998 to attack
$\ell_0$-minimization problems arising from signal and imaging
processing.

A large amount of empirical results (\cite{CDS98, CT05, CWB08,
BDE09, FL09, E10,  ZL12}) have shown that, in many situations,
$\ell_1$-minimization and its variants can locate  the sparsest
solution  to underdetermined linear systems. As a result, it is  important to rigorously address the fundamental
question: When does $\ell_1$-minimization solve
$\ell_0$-minimization? This question motivates one to identify the
conditions for the `equivalence' of these two problems. Let us first
clarify what we mean by `$\ell_0$- and $\ell_1$-minimization
problems are equivalent' in this paper.

\textbf{Definition 1.1.} (i)  \emph{ $\ell_0$- and
$\ell_1$-minimization  problems are said to be equivalent if  there
exists a solution to $\ell_0$-minimization  that coincides with the
unique solution to $\ell_1$-minimization.}  (ii) \emph{ $\ell_0$-
and $\ell_1$-minimization problems are said to be strongly
equivalent if the unique solution to $\ell_0$-minimization
coincides with the unique solution to $\ell_1$-minimization.}

Thus the \emph{equivalence} does not require that
$\ell_0$-minimization  have a unique  solution. In fact, an
underdetermined linear system may have multiple sparsest solutions.
Clearly, the \emph{strong equivalence} implies the
\emph{equivalence}, but the converse is not true. Currently, the
understanding of the relationship between $\ell_0$- and
$\ell_1$-minimization is mainly focused on the strong equivalence
(\cite{EB02, DE03, CT05, CRT06a, CRT06b, D06a, TD06, CWB08, BDE09,
SWM08, JN11, Z08}). The study of the strong equivalence is motivated by the newly
developed
 compressed-sensing theory (\cite{C06, D06}) which in
turn stimulates  the development of various sufficient criteria for
the strong equivalence between  $\ell_0$- and $\ell_1$-minimization.

The first work  on the strong equivalence (\cite{EB02, DE03, FN03, GN03,
F04, T04, MCW04}) claims that if a solution to the system $Ax=b$
satisfies that $\|x\|_0<\frac{1}{2}(1+\frac{1}{\mu(A)}),$ where
$\mu(A) $ is the mutual coherence of $A,$  then such a solution is
  unique   to both $\ell_0$- and  $\ell_1$-minimization.
The mutual-coherence-based analysis are very conservative. To go
beyond this  analysis, the restricted isometry property (RIP)
(\cite{CT05, C06}), null space property (NSP) (\cite{CDD09, Z08}),
ERC condition (\cite{T04, F04}), and other conditions (\cite{F04,
Z08, JN11}) were introduced over the past few years. The RIP and NSP
have been extensively investigated,  and   fruitful results on the
strong equivalence between $\ell_0$- and $\ell_1$-minimization  in
both noisy and noiseless data cases have been developed in the
literature. For instance, in noiseless cases, it has been shown by
E. Cand\`es that the RIP of order $2k$ implies that $\ell_0$- and
$\ell_1$-minimization are strongly equivalent whenever the linear
system has a solution satisfying $\|x\|_0 \leq k.$ The same
conclusion holds if certain NSP is satisfied (\cite{Z08, CDD09,
DDEK12}). However, RIP- and NSP-type conditions  remain restrictive,
compared to what the simulation has actually shown (\cite{CWB08,
E10}). Cand\`es and Romberg \cite{CR05} (see also \cite{CRT06a,
CRT06b}) initiated a probabilistic analysis for the efficiency of
the $\ell_1$-method. Following their work, various asymptotic and
probabilistic analyses of $\ell_1$-minimization have been
extensively carried out by many researchers (e.g., \cite{D06a, DT05,
CT05, E10,  PRT11}). The probabilistic analysis does demonstrate
that $\ell_1$-minimization has more capability of finding  sparse
solutions of linear systems than what is indicated by
state-of-the-art strong-equivalence sufficient criteria.

 In this paper, we introduce the so-called range space property
(RSP), a new matrix property governing the equivalence of $\ell_0$-
and $\ell_1$-minimization and the uniform recovery of sparse
signals. The RSP-based theory is motivated very naturally by the
needs of a further development of  the theory for sparse signal
recovery and
 practical applications.

  First, we note that the
existing sufficient criteria for the strong-equivalence of $\ell_0$- and
$\ell_1$-minimization are very restrictive. These criteria cannot
sufficiently interpret the actual numerical performance of the
$\ell_1$-method in many situations. They can only apply to a class
of linear systems with unique sparsest and unique least
$\ell_1$-norm solutions. In practice, the signal
 to recover may not be sparse enough, and the linear systems arising
from applications may have multiple sparsest solutions.  The
existing strong-equivalence-based theory fails in these situations.
(See Examples 3.4 and 3.6 in this paper.) For instance, when the
matrix $A$ is constructed by concatenating several matrices,  the
linear system often has multiple sparsest solutions. In this case,
we are still interested in finding a sparsest representation of the
signal $b$ in order for an efficient compression,  storing and
transmission of the signal. Our purpose in this case is to find one
sparsest solution of the underlying linear system. The
\emph{equivalence} of $\ell_0$- and $ \ell_1$-minimization can
guarantee to achieve this goal. The RSP turns out to be an
appropriate angle to approach the equivalence of $\ell_0$- and $
\ell_1$-minimization.

Second, from a mathematical point of view, there are  sufficient
reasons to develop the RSP-based theory.
 The mutual coherence
theory was developed from the Gram matrix $A^TA$ where $A$ has
normalized columns,
 the RIP of order $k$ is a property of the submatrices
$A_S^T A_S$ with $|S|\leq k$, and the NSP is a property of the null
space ${\cal N}(A). $  The RSP (introduced in this paper) is a
property of the range space ${\cal
 R}(A^T).$  Since $\ell_1$-minimization is a linear programming problem,  the
  optimality and/or the dual theory for this problem will unavoidably
lead to the RSP-based theory.

The RSP-based theory goes beyond the existing theory to guarantee
not only the strong equivalence but also the equivalence between
$\ell_0$- and $\ell_1$-minimization. It  applies  to a broader class
of linear systems than the existing theory, and it enables us to
  address the following questions: How to interpret the
numerical performance of $\ell_1$-minimization more efficiently than
the current theory?   If a linear system has multiple sparsest
solutions, when does $\ell_1$-minimization guarantee to find one of
them? How to deterministically interpret the efficiency and limit of
$\ell_1$-minimization for locating the sparsest solution to linear
systems? Can we further develop a theory for sparse signal recovery
 by using certain new matrix property rather than existing ones?

 To
 address these questions, our initial goal is to completely characterize the uniqueness of least
$\ell_1$-norm solutions to a linear system. It is the
classic strict complementarity theorem  of linear programming that
enables us to achieve this goal. Theorem 2.10 developed in this
paper claims that \emph{$x$ is the unique least $\ell_1$-norm
solution to the linear system $Ax=b$ if and only if the  so-called
range space property (RSP)   and a full-rank property hold at $x.$}

  Many
questions associated with the $\ell_1$-method, including the
above-mentioned ones,  can be largely addressed   from this theorem,
and a new theory for sparse signal recovery  can be   developed as well
(see sections 3 and 4 for details). For instance, the equivalence
 of $\ell_0$- and $\ell_1$-minimization
can be immediately obtained through this result. Our Theorem 3.3 in
this paper claims that \emph{a sparsest solution to a linear system
is the unique least $ \ell_1$-norm solution  to this system if and
only if it satisfies the range space property (irrespective of the
multiplicity of sparsest solutions)}. We note that the `if' part of
this result (i.e., the sufficient condition) was actually obtained
by Fuchs \cite{F04}. The `only if' part (i.e., the necessary
condition) is shown in the present paper. It is worth mentioning
that Donoho \cite{D05} has characterized the (strong) equivalence of
$\ell_0$- and $\ell_1$-minimization from a geometric (topological) point
of view, and Dossal \cite{D07} has shown that Donoho's geometric
result can be characterized by extending Fuchs' analysis.

The RSP-based analysis in this paper shows  that
 the uniqueness of
sparsest solutions  to linear systems is not necessary for the equivalence of $\ell_0$-
and $\ell_1$-minimization, and the multiplicity of sparsest solutions
may not prohibit the equivalence of these two problems as well. The
RSP is the only condition that determines whether or not a sparsest solution
to a linear system has a guaranteed recovery by
$\ell_1$-minimization. The RSP-based analysis can also explain the
numerical behavior of $\ell_1$-minimization more efficiently than
the strong-equivalence-based analysis.

 Moreover, we establish
several new recovery theorems, based on such matrix properties as
the RSP and Weak-RSP of order $K. $ Theorem 4.2 established in this
paper states that \emph{any $K$-sparse signal can be exactly
recovered by $\ell_1$-minimization if and only if $A^T$ has the RSP
of order $K.$} Thus the RSP of order $K$ can completely characterize
the uniform recovery of $K$-sparse signals. The Weak-RSP-based
recovery can be viewed as an extension of the  uniform
recovery. The key feature of this extended recovery theory is that
the uniqueness of sparsest solutions to   $Az=y $ may not be
required, where the vector $y$ denotes the measurements.

This paper is organized as follows.  A necessary and sufficient
condition for the uniqueness of least $\ell_1$-norm solutions to
linear systems is identified in section 2. The RSP-based equivalence
analysis for $\ell_0$- and $\ell_1$-minimization is given in section 3,
and
  the RSP-based recovery theory is developed in
section 4.

\emph{Notation.}
 Let $R^n$ be the $n$-dimensional Euclidean space,
and $R^n_{+}$  the first orthant of $R^n.$   For $x,y\in R^n$,
 $x\leq y$ means $x_i\leq y_i$ for every $i=1, ...,
n.$ Given a set $J\subseteq \{1,2,..., n\},$ the symbol $|J|$
denotes the cardinality of $J,$ and $J_c = \{1, 2,..., n\}
\backslash J $ is the complement of $J.$ For $x\in R^n,$
$\textrm{Supp} (x)= \{i: x_i\not=0\}$ denotes the support  of $x,$
$\|x\|_1=\sum_{i=1}^n |x_j| $ denotes the $\ell_1 $-norm of $x,$ and
$|x| = (|x_1|, ..., |x_n|)^T\in R^n $ stands for the absolute vector
of $x.$   Given a matrix $A=(a_1, ..., a_n)$ where $a_i$ denotes
$i$th column of the matrix, we use $A_S$ to denote a submatrix of
$A,$ with columns $a_i,$ $i\in S \subseteq \{1,2,..., n\},$ and
 $x_S $ stands for the subvector of $x\in R^n$ with components $x_i,$
$i\in S.$ Throughout the paper,  $e= (1,1,...,1)^T\in R^n$ denotes
the vector of ones.

\section{Uniqueness of least $\ell_1$-norm solutions}

   Let us first recall a classic theorem for
linear programming (LP) problems. Consider the LP problem
$$ (P)~~~~  \min \{c^T x: ~~ Qx=  p, ~~ x\geq 0\}, $$
and its dual problem
$$ (DP) ~~~~ \max\{p^Ty: Q^T y + s= c, ~ s\geq 0\},$$
where $Q$ is a given $m\times n$ matrix, and $p\in R^m$ and $  c\in
R^n$ are two given vectors.  Suppose that ($P$) and ($DP$) have
finite optimal values. By strong duality (optimality),
    $(x^*, (y^*, s^*))$ is a solution pair to the
 linear programming problems ($P$) and ($DP$) if and only if it
satisfies the  conditions $$
                             Qx^* =  p, ~~ x^*\geq 0,
                              ~Q^T y^* +s^* = c , ~ s^*\geq 0, ~
                             ~c^Tx^* = p^Ty^*,
                            $$
where $c^Tx^* = p^Ty^* $ can be equivalently written as $ x^*_i
s^*_i = 0 $ for every $ i=1, ..., n, $  which is called the
complementary slackness property. Moreover, the result below claims that when
($P$) and ($DP$) are feasible, there always exists a solution pair
$(x^*, (y^*, s^*))$ satisfying $x^*+s^*>0, $  which is called  a
strictly complementary solution pair.

 \textbf{Lemma 2.1} (\cite{S86}). \emph{Let (P) and (DP) be feasible. Then there exists a pair
 of strictly complementary  solutions to (P) and
(DP).}

We now develop some necessary conditions for $x$ to be the unique
least $\ell_1$-norm solution to the system   $Ax=b.$

\subsection{Necessary range property of $A^T$} Throughout this section, let $x$ be a given solution to the system $Ax=b.$
Note that for any other solution $z$ to this system, we have  that
$A(z-x)=0. $ So any solution $z$ can be represented as $z=x+u$ where
$u\in {\mathcal N}(A),$ the null space of $A.$ Thus we consider the
following two sets:
$$ C=\{u: \|u+x\|_1  \leq \|x\|_1\}, ~~  {\mathcal N}(A)=\{u: Au=0\}.$$
Clearly, $C$ depends on $(A,b,x).$ Since $(A,b,x)$ is assumed to be
given in this section, we use $C, $   instead of
$C(A,b,x), $ for simplicity

 First, we have the following straightforward observation, to which a simple proof is outlined in Appendix.

\textbf{Lemma 2.2. } \emph{The  following three statements are
equivalent:}

(i) \emph{ $x$ is  the unique least $\ell_1$-norm solution to the
system $Ax=b.$}

  (ii)  \emph{$C \bigcap {\mathcal N} (A) =\{0\}.$}

(iii)  \emph{$(u,t)= (0, |x|) $ is the unique solution to the system
  \begin{equation} \label {system}  Au=0, ~ \sum_{i=1}^n t_i\leq \|x\|_1,
  ~ |u_i+x_i| \leq t_i,  ~ i=1, ..., n.
   \end{equation}}

 It is evident that
$(u, t)=(0, |x|)$ is the unique solution to (\ref{system})  if and
only if  it is the unique solution to the LP problem
\begin{eqnarray*}   (LP_1)~~ & \min &   0^T  t \\
& \textrm{s.t.} &  Au=0,\\
  & &    \sum_{i=1}^n t_i\leq \|x\|_1,\\
              & &       |u_i+x_i|\leq t_i,  ~ i=1, ..., n ,
                    \end{eqnarray*}
  where $u$ and $ t$ are variables.
 % For every $i$ the
 % constraint  $|u_i+x_i|\leq t_i $  can be written as two inequalities:
 % $u_i+x_i-t_i\leq 0$ and $ -u_i-x_i-t_i \leq 0.$
 By introducing slack variables $\alpha, \beta\in R^n_+ $ and $ r \in R_+ , $
we can further write ($LP_1$)  as
\begin{eqnarray} (LP_2)~  &   \min & 0^T t  \nonumber \\
                  & \textrm{s.t.} & u_i+x_i  -  t_i +\alpha_i =0,   ~ i=1, ..., n,  \label{const1}\\
                    & & -u_i-x_i  -  t_i +\beta_i =0,   ~ i=1, ...,
                    \label{const2}
                    n, \\
 & & r+  \sum_{i=1}^n t_i  =  \|x\|_1, \nonumber \\
  & &  Au=0,  \nonumber\\
  & & \alpha \in R^n_+,  ~ \beta \in R^n_+, ~ r\geq 0, \nonumber
  \end{eqnarray}
  which is always  feasible, since $u=0, t=|x|, \alpha= |x|-x,
  \beta= |x|+x$ and $ r=0 $ satisfy all constraints. Almost all variables of ($LP_2$)
  are  nonnegative except $u.$ The nonnegativity of $t$ follows from (\ref{const1})
  and (\ref{const2}).   For the convenience of analysis,
  we  now transform ($LP_2$) into a form with all variables
  nonnegative.  Note that
 $u,$ satisfying  (\ref{const1}) and (\ref{const2}), is bounded. In fact,
  $$
  -2\|x\|_1   \leq  -x_i-t_i \leq u_i  \leq t_i-x_i   \leq 2\|x\|_1, ~ i=1, ..., n.
  $$
 Denote by  $M :=  2\|x\|_1+1 . $   Then $u'=Me-u \geq 0 $ for any $u$
 satisfying (\ref{const1}) and (\ref{const2}).
  Thus by the
  substitution $u=Me-u', $
    problem  ($LP_2$) can be finally written as
\begin{eqnarray*} (LP_3)~~~~ &  \min & 0^T t \\
                  & \textrm{s.t.} & (Me-u') -t +\alpha  =-x,   \\
                    & &  -(Me-u') -t +\beta  = x, \\
& & A(Me)-A u'=0,  \\
 & &  e^T t +r =  \|x\|_1, \\
 & &  (u',  t ,  \alpha, \beta, r)\in R^{4n+1}_+.
  \end{eqnarray*}
That is, {\small   \begin{eqnarray*}  & \min  &   0^T t \\
                  & \textrm{s.t.} &\\
                  & & \left[\begin{array}{lllll}
                                 -I & -I & I & 0 & 0 \\
                                 I & -I & 0 & I & 0 \\
                                 -A & 0 & 0 & 0 & 0 \\
                                 0 & e^T & 0 & 0 & 1 \\
                              \end{array}
                            \right]  \left[
                                       \begin{array}{c}
                                         u' \\
                                         t \\
                                         \alpha \\
                                         \beta \\
                                         r
                                       \end{array}
                                     \right]
                   = \left[
                                          \begin{array}{c}
                                            -x-Me \\
                                            x +Me\\
                                            -M (Ae) \\
                                            \|x\|_1 \\
                                          \end{array}
                                        \right],\\
                                       & &   (u',  t ,  \alpha, \beta, r)
                                        \geq 0. \end{eqnarray*} }
 From the above discussion, we have the next
 observation, to which the proof is given in Appendix.

\textbf{Lemma 2.3. }  \emph{The following three statements are
equivalent:}

 (i) \emph{$(u^*,t^*)= (0, |x|)$ is the unique solution to ($LP_1$).}

(ii) \emph{$(u^*, t^*, \alpha^*, \beta^*, r^*) =  (0, |x|, |x|-x,
|x|+x, 0) $ is the unique solution to (LP$_2).$}

(iii) \emph{$(u'^*, t^*, \alpha^*, \beta^*, r^*) =  (Me, |x|, |x|-x,
|x|+x, 0) $ is the unique solution to (LP$_3).$ }

 This lemma shows that
($LP_1$), ($LP_2)$ and ($LP_3$) are equivalent in the sense that if
one of them has a unique solution, so do the other two. Their unique
solutions are explicitly given in terms of $x. $   Note that the
dual problem of ($LP_3$) is given by
{\small \begin{eqnarray}  (DLP_3)   & \max &  -(x+Me)^T (y-y')- Me^TA^T y''+ \omega \|x\|_1 , \nonumber \\
                  & s.t. & -(y-y') -A^T y''\leq 0, \label{ineq-1} \\
                        & & -(y+y') +\omega e \leq 0, \label{ineq-2}\\
                        & & y\leq 0, \label{ineq-3} \\
                        & & y'\leq 0,  \label{ineq-4}\\
                        & & \omega \leq 0 \label{ineq-5}
  \end{eqnarray} }
  where $y, y', y'' $ and $\omega $ are the dual variables.
Let  $s^{(1)},$ $ s^{(2)},$ $ s^{(3)},$ $ s^{(4)} \in R^n_+$ and $ s
\in R_+^1 $ denote the nonnegative slack
 variables associated with the constraints (\ref{ineq-1}) through (\ref{ineq-5}), respectively, i.e.,
\begin{equation}\label{dualslack}  \left\{\begin{array}{l} s^{(1)} = (y-y') +A^T y'', \\    s^{(2)} = (y+y') -\omega e, \\
s^{(3)} =-y, \\ s^{(4)} =-y', \\   ~s=-\omega. \end{array} \right. \end{equation}
    We now prove that if $x$ is
    the unique least $\ell_1$-norm solution to the  system $Ax=b,$ then $ {\mathcal
    R}(A^T),$ the range space of $A^T,$  must satisfy certain property.

\textbf{Theorem 2.4.} \emph{If $x$ is the
 unique least $\ell_1$-norm solution to the system $Ax=b,$  then
there exist  $y, y'\in R^{n} $ and $\omega\in R$ satisfying
\begin{equation} \label{condition}  \left\{
\begin{array}{ll} y-y'\in {\mathcal R(}A^T), & \\
\omega < y_i+y'_i, ~y_i<0,  ~ y'_i<0   &  \textrm{ for }x_i=0,\\
y_i = 0 , ~  y'_i = \omega  &  \textrm{ for } x_i<0, \\
 y_i=\omega , ~ y'_i =0   & \textrm{ for } x_i>0.  \\
 \end{array}\right.
 \end{equation}}

   \emph{Proof.}  Assume that $x$ is the unique least $\ell_1$-norm solution to the system $Ax=b.$
  By Lemmas 2.2 and 2.3, the problem ($LP_3$) has a unique solution given by
  \begin{equation}\label{solution*}  (u'^*, t^*, \alpha^*, \beta^*, r^*)=(Me, ~|x|, ~|x|-x, ~|x|+x, ~0).
  \end{equation} Since both ($LP_3$) and its dual problem
  ($DLP_3$) are feasible, by Lemma 2.1, there exists a strictly complementary solution pair to ($LP_3$) and ($DLP_3$), denoted by
 $((\bar{u}',\bar{t},  \bar{\alpha}, \bar{\beta}, \bar{r}),
(y,y', y'', \omega) ). $    Since the solution to ($LP_3$) is
unique, we must have
\begin{equation} \label{sol***} (\bar{u}',\bar{t},  \bar{\alpha}, \bar{\beta},
\bar{r}) = (u'^*, t^*, \alpha^*, \beta^*, r^*).  \end{equation}  As
defined by (\ref{dualslack}), we use $(s^{(1)}, s^{(2)}, s^{(3)},
s^{(4)}, s)\in R^{4n+1}_+ $ to denote the  slack
 variables associated with  constraints  (\ref{ineq-1}) through (\ref{ineq-5})
 of  ($DLP_3$). By strict complementarity, we have
 \begin{equation} \label{compl-a} (\bar{u}')^Ts^{(1)} =0,  ~ \bar{t}^T s^{(2)} =0, ~ \bar{\alpha}^T
 s^{(3)} =0, ~\bar{\beta}^T s^{(4)}=0,   \bar{r} s=0, \end{equation}
 and
 \begin{equation} \label{compl-b} \bar{u}'+s^{(1)} > 0,  ~ \bar{t}+ s^{(2)} >0, ~
 \bar{\alpha}+
 s^{(3)} > 0,  \bar{\beta}+ s^{(4)} > 0,   \bar{r} + s > 0.   \end{equation}
 First, we see that $s^{(1)} = 0,$
 since $\bar{u}' = u'^*= Me
 >0.$ By the definition of $s^{(1)},$ it implies that
 \begin{equation} \label{constr-1} A^T y''= -(y-y').\end{equation}
 From (\ref{solution*}), we see that
$$ \begin{array}{rll}
t^*_i= x_i> 0,  & \alpha^*_i  =0,   &   \beta^*_i = 2x_i >0
\textrm{ for  }
x_i>0,\\
  t^*_i=|x_i|> 0,  & \alpha^*_i = 2|x_i| > 0 ,    & \beta^*_i = 0   ~ \textrm{ for  }  x_i< 0,\\
   t^*_i=0, &  \alpha^*_i =0, &  \beta^*_i  =  0   ~  \textrm{ for  }
i\not\in \textrm{Supp}(x),  \\
 r^*=0. &  &    \end{array}  $$ Therefore, it follows from (\ref{sol***}),
(\ref{compl-a}) and (\ref{compl-b})  that
$$\begin{array}{llll}
s^{(2)}_i= 0, & s^{(3)}_i> 0,  & s^{(4)}_i =0  & \textrm{  for  }
  x_i>0,\\
 s^{(2)}_i= 0, & s^{(3)}_i= 0,  & s^{(4)}_i > 0  & \textrm{  for
}   x_i<0,\\
 s^{(2)}_i >0, &  s^{(3)}_i> 0, &  s^{(4)}_i > 0  &  \textrm{  for
} i\not\in \textrm{Supp} (x),\\
  s>0.  & &  &
  \end{array} $$   By the
definition of these slack variables, the (strictly complementary)
solution vector
 $(y, y', y'', \omega)$ of
($DLP_3$)  satisfies (\ref{constr-1}) and $$ \begin{array} {llll}
\omega -(y_i+y'_i)=0,  & y_i<0, &   y'_i =0  & \textrm{ for } x_i>0,  \\
 \omega -(y_i+y'_i)=0, &  y_i=0,   & y'_i < 0 & \textrm{ for } x_i<0,  \\
 \omega - (y_i+y'_i) < 0, & y_i<0, &   y'_i <0  &  \textrm{ for } i\not\in
\textrm{Supp}(x),  \\
 \omega <0.  & & &
\end{array}  $$ Clearly,  the condition `$\omega<0$' is
redundant in the above system, since it is implied from other
conditions of the system.
 Thus we conclude that $(y,y',y'', \omega)$ satisfies (\ref{constr-1}) and the
following properties:
$$
\begin{array}{ll}
\omega < y_i+y'_i, ~y_i<0,  ~y'_i<0  &  \textrm{ for }x_i=0,\\
y_i = 0,  ~y'_i = \omega   & \textrm{ for } x_i<0, \\
 y_i=\omega, ~ y'_i =0  & \textrm{ for } x_i>0,\\
 \end{array}
 $$
which is exactly the condition (\ref{condition}), by noting that
  (\ref{constr-1}) is equivalent to  $y-y'\in {\mathcal
R} (A^T).$ ~~$\Box $

Therefore,   (\ref{condition})  is a necessary condition for $x$ to
be the unique least $\ell_1$-norm solution  to the system $Ax=b.$ This
condition arises naturally from the strict complementarity  of LP
problems. We now further point out that (\ref{condition}) can be
restated more concisely. The proof of this fact is given in
Appendix.

 \textbf{Lemma 2.5.}   \emph{Let  $x\in R^n$ be a given vector. There
exists a vector $(y, y',   \omega)\in R^{2n+1}$ satisfying
(\ref{condition}) if and only if there   exists a vector  $\eta \in
{\mathcal R}(A^T) $ satisfying that $ \eta_i = 1 $ for  all $
x_i>0,$
   $\eta_i =-1 $  for all $  x_i<0 , $  and
 $ |\eta_i| < 1$  for  all $ x_i =0.$}

Note that when
$\eta\in {\mathcal R}(A^T)$, both $-\eta $ and
 $\gamma \eta$ are also in ${\mathcal R}(A^T)$, where $\gamma$ is any real
 number. Thus the following three conditions are equivalent: (i) \emph{There is
an $\eta \in {\mathcal R}(A^T)$ with $
 \eta_i = 1$ for  $ x_i>0,$
  $ \eta_i =-1  $ for  $  x_i<0 , $  and
  $ |\eta_i| < 1 $  for $  x_i=0 .$ }
(ii)  \emph{There is  an $\eta \in {\mathcal R}(A^T) $ with $
 \eta_i = -1 $  for   $ x_i>0,$ $
    \eta_i =1 $  for  $ x_i<0 , $    and  $
  |\eta_i| < 1 $ for $  x_i=0 .$}
(iii) \emph{There is  an $\eta \in {\mathcal R}(A^T) $ with $
 \eta_i = \gamma $  for   $  x_i>0,
   \eta_i =-\gamma $  for  $  x_i<0 , $ and $
  |\eta_i| < |\gamma| $  for $  x_i=0 .$}
 The key feature here is that $\eta\in {\mathcal R }(A^T) $ has equal components
 (with value $\gamma\not=0$) corresponding to positive components of $x,$ and has equal
  components (with value $-\gamma$) corresponding to all negative components of $x,$ and
   absolute values of  other components of $\eta$ are strictly less that $|\gamma|.$
So if a linear system  has  a unique least $\ell_1$-norm solution,
the range space of  $A^T$ must satisfy the above-mentioned `nice'
property.

\subsection{Necessary full-rank condition}
In order to completely characterize the uniqueness of least
$\ell_1$-norm solutions to the system $Ax=b,$ we need to establish
another necessary condition. Let $x$ be a solution to the system
$Ax=b,$ and denote by $J_+=\{i: x_i>0\}$ and  $J_{-}=\{i: x_i<0\}. $
We have the following lemma.

 \textbf{Lemma 2.6.} \emph{The matrix \begin{equation}
\label{HHH} H =\left(
     \begin{array}{cc}
       A_{J_+} & A_{J_{-}}    \\
       -e^T_{J_+}  &  e^T_{J_{-}}     \\
     \end{array}
   \right)
\end{equation} has full column rank if and only if the
matrix below has full column rank \begin{equation} \label{GGG}
{\small G=\left(
     \begin{array}{cccc}
       -I_{|J_+|} & 0  & -I_{|J_+|} & 0 \\
       0 &  I_{|J_{-}|} & 0 & - I_{|J_{-}|}  \\
       A_{J_+} & A_{J_{-}} & 0 & 0  \\
       0  & 0 & e^T_{J_+} & e^T_{J_{-}} \\
     \end{array}
   \right), }
\end{equation}
where  $0$'s  are zero submatrices with suitable sizes, and
$I_{|J|}$ denotes the $|J|\times |J| $ identity matrix.}

 \emph{Proof. }
  Adding the first
$|J_+|+| J_{-}|$ rows of $G$ into its last row  yields the following
matrix $G'.$  Following that,  by similar column operations,   $G' $
can be further reduced to
   matrix $G'':$
{\small $$G'=\left[
     \begin{array}{cccc}
       -I_{|J_+|} & 0  & -I_{|J_+|} & 0 \\
       0 &  I_{|J_{-}|} & 0 & - I_{|J_{-}|}  \\
       A_{J_+} & A_{J_{-}} & 0 & 0  \\
       -e^T_{J_+}  &  e^T_{J_{-}}  &  0  & 0  \\
     \end{array}
   \right]  $$ $$  \rightarrow  G''=\left[
     \begin{array}{cccc}
       0 & 0  & -I_{|J_+|} & 0 \\
       0 &  0 & 0 & - I_{|J_{-}|}  \\
       A_{J_+} & A_{J_{-}} & 0 & 0  \\
       -e^T_{J_+}  &  e^T_{J_{-}}  &  0  &  0 \\
     \end{array}
   \right].
$$}
Note that upper-right block of $G''$ is a nonsingular square matrix,
and the lower-left block is $H.$ Since any elementary row and column
operations do not change the (column) rank of a matrix. Thus $H$ has
full column rank if and only if   $G $ has full column rank. ~~
$\Box$

We now prove the next necessary condition for the uniqueness of
least $\ell_1$-norm solutions to the system $Ax=b.$

 \textbf{Theorem 2.7.} \emph{If $x$ is the unique least $\ell_1$-norm
 solution to the system $Ax=b,$ then the matrix $H,$ defined by
 (\ref{HHH}), has full column rank.}

\emph{Proof.}   Assume the contrary that the columns of $H$ are
linearly dependent. Then by Lemma 2.6, the columns of matrix $G,$
given by (\ref{GGG}), are also linearly dependent. Hence, there
exists a vector $d=(d_1, d_2, d_3, d_4) \not=0 ,$ where $d_1, d_3\in
R^{|J_+|}$ and $ d_2, d_4\in R^{|J_{-}|},$ such that {\small
$$ G d= \left(
     \begin{array}{cccc}
       -I_{|J_+|} & 0  & -I_{|J_+|} & 0 \\
       0 &  I_{|J_{-}|} & 0 & - I_{|J_{-}|}  \\
       A_{J_+} & A_{J_{-}} & 0 & 0  \\
       0  & 0 & e^T_{J_+} & e^T_{J_{-}} \\
     \end{array}
   \right) \left(
             \begin{array}{c}
               d_1 \\
               d_2 \\
               d_3 \\
               d_4 \\
             \end{array}
           \right)
    = 0.$$ }
Note that $ z=(z_1, z_2, z_3,z_4),$ where
\begin{equation}\label{zzzz} \left\{\begin{array}{l} z_1= Me_{J_+}>0, \\ z_2 = Me_{J_{-}}>0,\\
  z_3 = x_{J_+} >0,  \\ z_4  = -x_{J_{-}}>0, \end{array} \right. \end{equation}
is a solution to   the  system
 \begin{eqnarray}\label{SSMM}  & &   \left(
     \begin{array}{cccc}
       -I_{|J_+|} & 0  & -I_{|J_+|} & 0 \\
       0 &  I_{|J_{-}|} & 0 & - I_{|J_{-}|}  \\
       A_{J_+} & A_{J_{-}} & 0 & 0  \\
       0  & 0 & e^T_{J_+} & e^T_{J_{-}} \\
     \end{array}
   \right) \left(
             \begin{array}{c}
               z_1 \\
               z_2 \\
               z_3 \\
               z_4 \\
             \end{array}
           \right)  \nonumber \\
           & &  = \left(
             \begin{array}{c}
                -x_{J_+}- Me_{J_+} \\
                x_{J_{-}}+Me_{J_{-}} \\
                M(A_{J_+} e_{J_+} + A_{J_{-}}e_{J_{-}})  \\
                 \|x_{J_+}\|_1 + \|x_{J_{-}}\|_1 \\
             \end{array}
           \right).    \end{eqnarray}
Since $z>0,$ there exists a small number $\lambda \not =0$ such that
$\widetilde{z}=(\widetilde{z}_1, \widetilde{z}_2, \widetilde{z}_3,
\widetilde{z}_4) = z+\lambda d \geq 0$ is also a (nonnegative)
solution to the system (\ref{SSMM}), where $\widetilde{z}_i=
z_i+\lambda d_i, i=1, ..., 4. $  Clearly, $\widetilde{z} \not = z $
since $\lambda \not=0$ and $ d \not =0.$ We now construct two
solutions to ($LP_3$) as follows. Let
$$ u'= \left[
         \begin{array}{c}
           z_1 \\
           z_2 \\
           Me_{J_0} \\
         \end{array}
       \right] \in R^n, t= \left[
         \begin{array}{c}
           z_3 \\
           z_4 \\
           0 \\
         \end{array}
       \right] \in R^n,  $$ $$ \alpha = \left[
         \begin{array}{c}
           0 \\
           2z_4 \\
           0 \\
         \end{array}
       \right] \in R^n , \beta= \left[
         \begin{array}{c}
           2z_3 \\
           0 \\
           0 \\
         \end{array}
       \right]\in R^n, $$ and $r=0, $
where $z_1, z_2, z_3$ and $ z_4$ are given by (\ref{zzzz}), and
$J_0=\{i: x_i=0\}.$  It is not difficult to see that the vector
$(u',t,\alpha, \beta, r)$  defined above is exactly the one
$$ (u'=Me, ~ t=|x|, ~ \alpha= |x|-x, ~ \beta=|x|+x, ~ r=0). $$ On the other
hand, we can also define
$$  \widetilde{u}'= \left[
         \begin{array}{c}
           \widetilde{z}_1 \\
           \widetilde{z}_2 \\
           Me_{J_0} \\
         \end{array}
       \right] \in R^n, \widetilde{t}= \left[
         \begin{array}{c}
           \widetilde{z}_3 \\
           \widetilde{z}_4 \\
           0 \\
         \end{array}
       \right] \in R^n, $$ $$ \widetilde{\alpha} = \left[
         \begin{array}{c}
           0 \\
           2\widetilde{z}_4 \\
           0 \\
         \end{array}
       \right] \in R^n , \widetilde{\beta}= \left[
         \begin{array}{c}
           2\widetilde{z}_3 \\
           0 \\
           0 \\
         \end{array}
       \right]\in R^n,$$ and  $\widetilde{r}=0. $
Clearly, $(\widetilde{u}', \widetilde{t}, \widetilde{\alpha},
\widetilde{\beta}, \widetilde{r})$ is in $ R^{4n+1}_+,$ and by a
straightforward verification, we can show that this vector satisfies
all constraints of $(LP_3). $ So $ (\widetilde{u}', \widetilde{t},
\widetilde{\alpha}, \widetilde{\beta}, \widetilde{r})$ is also a
solution to $(LP_3).$ This implies that
  ($LP_3$) has two different
solutions: $(u', t, \alpha, \beta, r)\not = (\widetilde{u}',
\widetilde{t}, \widetilde{\alpha}, \widetilde{\beta},
\widetilde{r}). $ Under the assumption of the theorem, however, it
follows from Lemmas 2.2 and 2.3 that ($LP_3$) has a unique solution.
This contradiction shows that $H$, defined by (\ref{HHH}), must have
full column rank. ~~ $\Box$

Combining  Lemma 2.5, Theorems 2.4  and  2.7 yields the next result.

 \textbf{Theorem 2.8.}  \emph{If $x$ is the unique least
$\ell_1$-norm solution to the system  $Ax=b,$ then (i) the matrix $
\left(
     \begin{array}{cc}
       A_{J_+} & A_{J_{-}}    \\
       -e^T_{J_+}  &  e^T_{J_{-}}     \\
     \end{array}
   \right)
$ has full column rank, and (ii) there exists a vector  $\eta  $
such that
\begin{equation}\label{range} \left\{\begin{array} {ccccc} \eta  &\in & {\mathcal
R}(A^T), & &  \\
 \eta_i &  =  & 1 &  \textrm{ for all }  & x_i>0, \\
 \eta_i  & = & -1  &  \textrm{ for all }  & x_i<0 , \\
|\eta_i|  & < & 1 & \textrm{ for all }  & x_i =0.
\end{array}\right. \end{equation}}

 In this paper, the   condition (ii) above
  is called the \emph{range space property} (RSP) of $A^T$ at $x.$
  It is worth noting that checking the RSP of $A^T$ at $x$ is very easy. It is equivalent to solving
 the LP problem
 \begin{eqnarray*} &  \min & \tau \\
                  & \textrm{s.t}. &   A_{J_+}^T y= e_{J_+}, \\
                  & &   A_{J_-}^T y= - e_{J_-}, \\
 & & A_{J_0}^Ty = \eta_{J_0},  ~  |\eta_{J_0}| \leq \tau e_{J_0},
\end{eqnarray*}
where $J_0=\{i: x_i=0\}.$
 Clearly, the RSP of $A^T$ at $x$ holds if and only if the
optimal value of the LP problem above satisfies $\tau <1.$

 \subsection{A necessary and sufficient condition}

Clearly, if (\ref{range}) holds,  there is a vector $u$ such that
$\left(
 \begin{array}{c}
  e_{J_+}  \\
    -e_{J_{-}} \\
     \end{array}
      \right)$ $ = \left(
  \begin{array}{c}
      A^T_{J_+} \\
A^T_{J_{-}}
  \end{array}
   \right) u ,$  and thus under condition (\ref{range}), we have
   $\textrm{rank} \left(
     \begin{array}{cc}
       A_{J_+} & A_{J_{-}}    \\
       -e^T_{J_+}  &  e^T_{J_{-}}     \\
     \end{array}
   \right)  = \textrm{rank} (
A_{J_+}  ~ A_{J_{-}}
   ). $ Thus condition (i) in Theorem 2.8 can be
   simplified to that the matrix $ (
       A_{J_+} ~ A_{J_{-}})
$ (i.e., $A_{Supp(x)})$ has full column rank.

As we have seen from the above analysis, it is the  strict
complementarity theory of linear programming that leads to the
necessary conditions  in Theorem 2.8 which is established for the
first time in this paper. The strict complementarity theory can be
also used to prove that the converse of Theorem 2.8 is true, i.e.,
(i) together with (ii) in Theorem 2.8 is also sufficient for
$\ell_1$-minimization to have a unique solution. However, we omit
this proof since this sufficiency was obtained already by Fuchs
\cite{F04}, while his analysis was based on convex quadratic
optimization, instead of strict complementarity.

\textbf{Theorem 2.9} (Fuchs \cite{F04}).  \emph{If the system $Ax=b$
has a solution $x$ satisfying   (\ref{range}) and the columns of $A$
corresponding to the support of $x$ are linearly independent, then
$x$ is the unique least $\ell_1$-norm solution to the system $Ax=b.
$}

The necessary condition (Theorem 2.8) developed in this paper is the
key to interpret the efficiency and limit of $\ell_1$-minimization
in locating/recovering sparse signals (see the discussion in the
next section). Each of Theorems 2.8 and 2.9 alone can only give a
half picture of the uniqueness of the least $\ell_1$-norm solutions
to a linear system. Theorems 2.8 and 2.9 together yield the
following complete characterization.

\textbf{Theorem 2.10.} \emph{$x$ is the unique least $\ell_1$-norm
solution to the system $Ax=b $ if and only if the RSP (\ref{range})
holds at $x,$ and the matrix $(A_{J_+} ~ A_{J_{-}}) $ has full
column rank,  where $ J_+= \{i: x_i>0 \}$ and $ J_{-}= \{i:
x_i<0\}.$  }

This necessary and sufficient condition provides a good basis to
interpret the relationship of $\ell_0$- and $\ell_1$-minimization,
the internal mechanism of the $\ell_1$-method, and their roles in
compressed sensing.

\section{RSP-based analysis for the equivalence of $\ell_0$- and $\ell_1$-minimization}

    In this
section, we focus on the condition for the equivalence
  of $\ell_0$- and $\ell_1$-minimization. Through this and the next
  sections, we will see how Theorem 2.10 can be used to interpret the efficiency  and limit
  of the $\ell_1$-method in finding/recovering sparse solutions of
  linear systems. For the convenience of discussion, we clarify
   the   concept below.

  \textbf{Definition 3.1.}  \emph{A solution $x$ to the system $Ax=b$ is said to
  have a guaranteed recovery (or to be exactly recovered) by the $\ell_1$-method   if $x$ is the
  unique least $\ell_1$-norm solution to this system.}

Note that when  matrix $(A_{J_+} ~ A_{J_{-}})$  has full column
rank, the number of its columns (i.e., $|J_+|+|J_{-}| =\|x\|_0$) is
less than or equal to the number of its rows.  Therefore, the
following fact is
  implied from Theorem 2.10.

\textbf{Corollary 3.2.} \emph{If $x$ is the unique least
$\ell_1$-norm solution to the system $Ax=b$, where $A\in R^{m\times
n}$ with $m<n,$ then $\|x\|_0 \leq m.$}

This result shows that when $\ell_1$-minimization has a unique
solution, this solution must be at least $m$-sparse. In other words,
any $x\in R^n$ that can be exactly recovered by
$\ell_1$-minimization must be a sparse vector with $\|x\|_0 \leq m.$
This property of the $\ell_1$-method justifies its role as a
sparsity-seeking method. Corollary 3.2 also implies that any  $x$
with sparsity $\|x\|_0
> m$ is definitely not the unique least $\ell_1$-norm solution to a linear system,
and hence there is no guaranteed recovery for such a solution  by
  $\ell_1$-minimization. Note that Gaussian elimination or other
linear algebra methods can also easily find a solution  with
$\|x\|_0 \leq m , $ although there is no guarantee for $\|x\|_0 <
m.$ Thus what we really want from the $\ell_1$-method is a truly
sparse solution with $\|x\|_0 < m$ if such a solution exists. So it
is natural to ask when the $\ell_1$-method finds a sparsest
solution. By Theorem 2.10, we have the following result that
completely characterizes the equivalence of $\ell_0$- and
$\ell_1$-minimization.

 \textbf{Theorem 3.3.} \emph{Let $x\in R^n$ be a
sparsest solution to the system  $Ax=b.$ Then $x$ is the unique
$\ell_1$-norm solution to this system  if and only if the range
space property defined by (\ref{range})  holds at $x.$}

\emph{Proof.} On one hand, Theorem 2.10  claims that when a solution
$x$   is the unique least $\ell_1$-norm solution, the RSP
(\ref{range}) must be satisfied at this solution. On the other hand,
when $x$ is the sparsest solution to $Ax=b$, the column vectors of
$A$ corresponding to the support  of $x$ must be linearly
independent (i.e., $A_{Supp(x)} $ has  full column rank), since
otherwise at least one of the  columns of $A_{Supp(x)} $ can be
represented by other columns, and hence a solution sparser than $x$
can be found, leading to a contradiction. So the matrix $(A_{J_+} ~
A_{J_{-}})$  always has full column rank at any sparsest solution
$x$ of the system  $Ax=b.$  Thus by Theorem 2.10 again, to guarantee
a sparsest solution to be the unique least $\ell_1$-norm solution,
the only condition required is the RSP.  The desired result follows.
~~ $\Box$

Although Theorem 3.3 is a special case of Theorem 2.10,
 it is powerful enough to encompass all existing sufficient conditions
 for the  strong equivalence of $\ell_0$- and
 $\ell_1$-minimization as special cases,
  and it goes  beyond the scope of these conditions.
     To show this, let us first
      decompose the underdetermined linear systems  into
      three categories as follows:

\begin{itemize}
 \item  \emph{Group 1}: The system has a \emph{unique} least $\ell_1$-norm solution and a \emph{unique} sparsest solution.

  \item \emph{Group 2}: The system has a \emph{unique} least $\ell_1$-norm
solution and  \emph{multiple} sparsest solutions.

  \item \emph{Group 3}: The system has \emph{multiple} least $\ell_1$-norm solutions.
\end{itemize}

Clearly, every linear system falls into one and only one of
      these groups. By Theorem 2.10,   the linear systems in Group
      3
       do not satisfy either the RSP  or full-rank property at any of its
       solutions. Thus the guaranteed recovery by $\ell_1$-minimization can  only possibly happen
 within Groups 1 and 2.
Since many existing sufficient conditions for the equivalence of
$\ell_0$- and $ \ell_1$-minimization actually imply the strong
equivalence between these two problems,  these
conditions can only  apply to  a subclass of linear systems in Group
1. The following three conditions are widely used in the literature:

\begin{itemize} \item   Mutual coherence condition (\cite{DE03, GN03, EB02, FN03}) $
\|x\|_0 \leq \frac{1}{2}(1+\frac{1}{\mu(A)}),$ where $\mu(A)$ is
defined by $\mu(A)=\max_{1\leq i,j\leq m, i\not = j} |a^T_i
a_j|/(\|a_i\|_2 \cdot \|a_j\|_2).  $

\item  Restricted isometry property (RIP) \cite{CT05}. The matrix $A$
has the restricted isometry property (RIP) of order $k$ if there
exists a constant $0<\delta_k<1 $ such that
$(1-\delta_k)\|z\|_2^2\leq \|Az\|_2^2\leq (1+\delta_k)\|z\|_2^2$ for
all $k$-sparse vector $z.$

\item  Null space property (NSP) (\cite{CDD09, DDEK12}).   The matrix $A$
has the NSP of order $k$ if there exists a constant $\vartheta
 >0 $ such that  $ \|h_\Lambda \|_2 \leq \vartheta
\frac{\|h_{\Lambda_c}\|_1}{\sqrt{k} }$ holds for all $h\in {\mathcal
N} (A) $ and  all $\Lambda \subseteq\{1, 2, ..., n\}$ such that
$|\Lambda|\leq k.$ The NSP can be defined in other ways (see
e.g., \cite{Z08}).
\end{itemize}
Some other important conditions can be also found in the literature, such as
the accumulative coherence condition (\cite{T04, DE03}),  exact
recovery coefficient (ERC) \cite{T04}, and others (see e.g.,
 \cite{Z08,JN11}).  Theorem 3.3 shows that
 \emph{all existing
conditions for the   equivalence between $\ell_0$- and $\ell_1$-minimization
in the literature imply the RSP (\ref{range}),} since the RSP is not
only a  sufficient, but also a necessary condition for the
equivalence between $\ell_0$- and $\ell_1$-minimization problems.

A remarkable difference between the RSP and many existing sufficient
conditions is that the RSP does not require the uniqueness of
sparsest solutions.
 Even if  a linear system has multiple sparsest solutions, $\ell_1$-minimization
 can still guarantee to solve an
$\ell_0$-minimization problem, provided that  the RSP holds at a sparsest
solution of the problem.

\textbf{Example 3.4.} (RSP \emph{does not require the uniqueness of
 sparsest solutions}). Consider the linear system   $Ax=b
$  with {\small $$ A = \left(
  \begin{array}{ccrr}
    1 & 0 & -2 & 5 \\
    0 & 1 & 4 & -9 \\
    1 & 0 & -2 & 5 \\
  \end{array}
\right),\textrm{ and } b= \left(
          \begin{array}{r}
            1 \\
            -1 \\
            1 \\
          \end{array}
        \right). $$ }
It is easy to see that the system $Ax=b$ has multiple sparsest
solutions: $x^{(1)} =(1, -1, 0, 0)^T,$ $ x^{(2)}  =(0, 1, -1/2,
0)^T,$ $ x^{(3)}= (0, 4/5, 0, 1/5)^T, $ $ x^{(4)} =(0, 0, 2, 1)^T, $
$x^{(5)} = (1/2, 0, -1/4, 0)^T$ and $ x^{(6)}= (4/9, 0, 0, 1/9)^T.$
We now verify that the RSP holds at $x^{(6)}.$ It is sufficient to
find a vector $\eta =(1, \eta_2, \eta_3, 1)$ in the range space of
$A^T$ with $|\eta_2|<1$ and $|\eta_3|<1.$ Indeed, by taking $u=(1,
4/9, 0)^T, $ we have that $\eta = A^T u = (1, 4/9, -2/9, 1)^T. $
Thus the RSP holds at $ x^{(6)} $ which, by Theorem 3.3, has a
guaranteed recovery by $\ell_1$-minimization. So $\ell_0$- and
$\ell_1$-minimization problems are equivalent. It is worth noting that the mutual
coherence, RIP and NSP  cannot apply to this example, since the
system has multiple sparsest solutions.

  It is easy to check that among all 6 sparsest solutions of the above example,  $x^{(6)}$ is the
only sparsest one satisfying the RSP. This is not a surprise, since
by Theorem 3.3 any sparsest solution  satisfying the RSP must be  the
unique least $\ell_1$-norm solution. Thus we have the following
corollary.

\textbf{Corollary 3.5.} \emph{For any given underdetermined linear
system, there exists at most one sparsest solution satisfying RSP
(\ref{range}).}

 The next  example shows that even for a
problem  in Group 1,   many existing sufficient conditions may fail
to confirm the strong equivalence of $\ell_1$- and
$\ell_0$-minimization, but the RSP can successfully confirm this.

 \textbf{Example 3.6.}  Consider the system $Ax=b$ with
$$ A = \left(
  \begin{array}{cccrrr}
    \frac{1}{\sqrt{2}} & 0 & \frac{1}{\sqrt{3}} & -\frac{1}{\sqrt{2}}  & \frac{1}{\sqrt{2}} & 0 \\
     \frac{1}{\sqrt{2}} &  \frac{1}{\sqrt{2}}  &  \frac{1}{\sqrt{3}} & - \frac{1}{\sqrt{3}} & 0 & -\frac{1}{\sqrt{2}}\\
     0  & \frac{1}{\sqrt{2}}  &  \frac{1}{\sqrt{3}} & \frac{1}{\sqrt{6}} & -\frac{1}{\sqrt{2}} & -\frac{1}{\sqrt{2}} \\
  \end{array}
\right) ,  b= \left(
                \begin{array}{c}
                  1 \\
                  1 \\
                  1 \\
                \end{array}
              \right)
  $$
Clearly, $x^*= (0, 0, \sqrt{3},0, 0,0) $ is the unique sparsest
solution to this linear system. Note that  $a_2^T a_6 =-1$ which
implies that $\mu(A)= 1$. The mutual coherence condition $\|x^*\|_0
< \frac{1}{2} (1+\frac{1}{\mu(A)})=1$ fails. Since the second and
last columns are linearly dependent, the RIP of order $2$ fails.
Note that $\eta=(0,1,0,0,0,1)^T\in {\mathcal N}(A)$ does not satisfy
the null space property of order $2.$ So the NSP of order 2 also
fails.
  However, we can find a   vector
$\eta = (\eta_1, \eta_2, 1, \eta_4, \eta_5, \eta_6)$ in ${\mathcal
R}(A^T)$ with $|\eta_i|<1$ for all $i\not =3.$ In fact, by simply
taking $y= (\frac{1}{\sqrt{3}}, \frac{1}{\sqrt{3}},
\frac{1}{\sqrt{3}})$, we have
$$ \eta = A^Ty =  \left(\sqrt{\frac{2}{3}}, \sqrt{\frac{2}{3}}, 1,
\frac{1-\sqrt{3}-\sqrt{2}}{3\sqrt{2}}, 0, -
\sqrt{\frac{2}{3}}\right). $$  Thus the RSP (\ref{range})   holds at
$x^*.$ By Theorem 3.3, $\ell_0$- and $\ell_1$-minimization are
equivalent, and thus $x^*$ has a guaranteed recovery by
$\ell_1$-minimization.

 From the above discussion, we have actually shown, by Theorem 3.3, that the equivalence between
 $\ell_0$- and $ \ell_1$-minimization
  can be achieved not only for a subclass of problems in Group 1, but also for a subclass of problems
  in large Group 2. Since many existing sufficient conditions imply the strong
equivalence between $\ell_0$- and $\ell_1$-minimization  which can
be achieved only for a subclass of problems in Group 1, these
conditions cannot apply to  linear systems in Group 2, and hence
cannot explain the success of $\ell_1$-minimization for solving
$\ell_0$-minimization with multiple sparsest solutions.  The
simulation shows that the $\ell_1$-method performs much better than
what has predicted by those strong-equivalence-type conditions. Such
a gap between the current theory and the actual performance of the
$\ell_1$-method can be clearly interpreted and identified now by our
RSP-based analysis. This analysis indicates that the uniqueness of
sparsest solutions is not necessary for an $\ell_0$-minimization to
be solved by the $\ell_1$-method, and the multiplicity of sparsest
solutions of a linear system does not prohibit the $\ell_1$-method
from solving an $\ell_0$-minimization as well.  When many existing
sufficient conditions fails (as shown by Examples 3.4 and 3.6), the
RSP-based analysis shows that the $\ell_1$-method can continue its
success in solving $\ell_0$-minimization problems in many
situations. Thus  it does show that the actual success rate of
$\ell_1$-minimization for solving $\ell_0$-minimization problems is
certainly higher than what has indicated by the
strong-equivalence-based theory.
 Moreover, the RSP-based theory
also sheds light on the limit  of  $\ell_1$-minimization. Failing to
satisfy the RSP, by Theorem 3.3 a sparsest solution is definitely
not the unique least $\ell_1$-norm solution, and hence there is no
guaranteed recovery for such a solution by $\ell_1$-minimization.

 \section{Compressed sensing: RSP-based sparsity recovery}

 So far, the sparsity recovery theory has been developed by various approaches,
 including the $\ell_1$-method (i.e., the so-called basis pursuit) and heuristic methods
 such as the (orthogonal)
 matching pursuit (e.g. \cite{TG07, E10}).  In this section,
 we show how Theorems 2.10 and 3.3 can be used to develop  recovery criteria for sparse signals.
 Suppose that we would like to recover a sparse
vector $x^*.$ To serve this purpose, the so-called sensing matrix
$A\in R^{m\times n}$ with $m<n$ is constructed, and the measurements
$y=Ax^*$ are taken. Then we solve the $\ell_1$-minimization problem
$\min\{\|x\|_1: Ax=y\} $ to obtain a solution $\widehat{x}.$  The
compressed sensing theory is devoted to addressing, among others,
the following questions: What class of sensing matrices can
guarantee the exact recovery $ \widehat{x}= x^*$, and how sparse
should $x^*$ be in order to achieve the recovery success?  To
guarantee an exact recovery, $A$ is constructed to satisfy the
following three conditions:

\begin{itemize}
\item[(C1)] $Ax=y$ has a unique least $\ell_1$-norm solution
$\widehat{x}.$

\item[(C2)]$x^*$ is the unique sparsest
solution to  the system $Ax=y . $

\item[(C3)]  These two  solutions are equal.
\end{itemize}
Clearly, satisfying (C1)--(C3) actually requires that $\ell_0$- and
$ \ell_1$-minimization are strongly equivalent. Many existing
recovery theories comply with this framework.  For instance, if $A$
satisfies the RIP of order $2k$ with $\delta_{2k} < \sqrt{2}-1$ (see
\cite{C06}), or if $A$ satisfies the NSP of order $2k$ (see
\cite{CDD09, DDEK12}), then  there exists a constant $ \vartheta$
such that for any $x\in R^n$, it holds
\begin{equation} \label{estimate} \|x-x^*\|_2\leq \vartheta
\sigma_k(x)/\sqrt{k},
\end{equation}
where $ \sigma_k(x)=\min \{\|x-z\|_1: \|z\|_0\leq k\},$  the
$\ell_1$-norm of the $n-k$ smallest components of $x.$ This result
implies that if $Ax=y $ has a solution satisfying $\|x\|_0\leq k,$
  it must be equal to $x^*, $ and it  is   the
  unique sparsest solution to the system $Ax=y= Ax^*. $

\subsection{RSP-based uniform recovery}
Recall that the spark of a given matrix, denoted by
$\textrm{spark}(A),$  is the smallest number of columns of A that
are linearly dependent \cite{DE03}. The exact recovery of all
$k$-sparse vectors (i.e., $\{x: \|x\|_0\leq k\}$) by a single
sensing matrix $A$ is called \emph{uniform recovery}. It is well
known that the RIP and NSP of order $2k$ can uniformly recover
$k$-sparse vectors, where $k<\frac{1}{2}\textrm{spark}(A). $ We now
characterize the uniform recovery by a new concept defined as
follows.

\textbf{Definition 4.1.} (RSP of order $K$)  \emph{Let $A\in
R^{m\times n}$ with $m<n.$  The matrix $A^T$ is said to satisfy the
range space property of order $K$ if for any disjoint subsets $S_1,
S_2$ of $\{1,..., n\}$ with $|S_1|+|S_2|\leq
 K$, the range space $ {\mathcal R}(A^T) $ contains a vector $\eta$ such
 that $\eta_i =1 $ for all $ i \in S_1,$  $ \eta_i =-1
 $ for all $ i\in
 S_2, $ and $ |\eta_i| <1 $ for all other components.}

This concept can be used to   characterize the uniform
recovery, as shown by the next result.

\textbf{Theorem 4.2. } (i)  \emph{If $A^T$ has the RSP of order $K$,
then  any $K$ columns of $A$ are linearly independent. }  (ii)
\emph{Assume that the measurements of the form $y= Ax$ are taken.
Then  any $x$ with $\|x\|_0 \leq K$ can be exactly recovered by
$\ell_1$-minimization  if and only if $A^T$ has the RSP of order
$K.$ }

\emph{Proof.} (i) Let $S\subseteq \{1,..., n\}$ be any subset with
$|S|=K. $ We denote the elements of $S$   by $\{s_1, ..., s_K\}.$ We
prove that the columns of $A_S$ are linearly independent.
  It is sufficient to show that $z_S=0$ is the only
solution to $A_Sz_S=0. $  In fact, assume  $A_Sz_S=0. $  Then
$z=(z_S, z_{S_c} =0 ) \in R^n $ is in the null space of $A. $
Consider the disjoint sets $S_1=S,$ and $S_2 =\emptyset . $ By the
RSP of order $K$, there exists a vector $\eta \in {\mathcal R}(A^T)
$ with $\eta_i=1$ for all $i\in S_1 =S.$ By the orthogonality of
${\cal N}(A)$ and ${\cal R}(A^T),$ we have $0= z^T \eta= z_S^T
\eta_S+z_{S_c}^T \eta_{S_c} =z_S^T \eta_S,$ i.e.,
\begin{equation} \label{zzss} \sum_{j=1}^K  z_{s_j} =0. \end{equation}  Now we consider any $k$ with
$1\leq k\leq K$, and the pair of disjoint sets:  $$S_1=\{s_1, s_2,
..., s_k\}, ~~ S_2=\{s_{k+1}, ..., s_K\}. $$ By the RSP of order
$K$, there exists an $\eta\in {\mathcal R}(A^T) $ with
$\eta_{s_i}=1$ for every $i=1,..., k $ and $\eta_{s_i}=-1$  for
every  $i=k+1, ..., K.$ By  orthogonality again,  it follows from
$z^T\eta=0$   that
$$ (z_{s_1}+ \cdots +z_{s_k}) - ( z_{s_{k+1}}+ \cdots + z_K) =0,  $$ which holds for every $k$
with $1\leq k\leq K. $ It follows from these relations, together
with (\ref{zzss}),  that all components of $z_S$ must be zero. This
implies that any $K$ columns of $A$ are linearly independent.

(ii) First we assume that the RSP of order $K$ is satisfied.  Let
$x$ be an arbitrary vector with $\|x\|_0 \leq K.$ Let $S_1=J_+= \{i:
x_i>0\}$ and $S_2=J_{-}= \{i: x_i<0\}.$ Clearly, $S_1$ and $S_2$ are
disjoint, and $|S_1|+|S_2|\leq K,$ by the RSP of order $K$, there
exists a vector $\eta \in {\mathcal R} (A^T) $ such that $\eta_i =1
$ for all $ i \in S_1,$  $ \eta_i =-1
 $ for all $ i\in
 S_2, $ and $ |\eta_i| <1 $ for all other components. This implies
  that the  RSP  (\ref{range}) holds at $x$.
 Also, it follows from (i) that any $K$ columns of $A$ are linearly independent, and thus any
 $|S_1|+|S_2| ~(\leq K) $ columns of $A$ are
 linearly independent. So
 the matrix $ (A_{S_1}  ~ A_{S_2})$ has full column rank. By Theorem 3.3 (or 2.10),
$x$ is the unique least $\ell_1$-norm solution
   to the  equation $Az=y.$ So $x$ can be exactly recovered by $\ell_1$-minimization.

Conversely,  assume that any $K$-sparse vector can be exactly
recovered by $\ell_1$-minimization. We prove that $A^T$ satisfies
the RSP of order $K.$ Indeed, let $x$ be a $K$-sparse vector, and
let $y$ be the measurements, i.e., $y=Ax.$ Under the assumption, $x$
can be exactly recovered by the $\ell_1$-method, so $x$ is the
unique least $\ell_1$-norm solution to the system $Az=y.$ By Theorem
2.8 (or 2.10), the RSP (\ref{range}) holds at $x. $ This implies
that there exists $\eta\in {\mathcal R}(A^T)$ such that $\eta_i=1$
for $i\in S_1,$ $ \eta_i=-1 $ for $ i\in S_2 , $ and $|\eta_i |<1 $
for all other components, where $S_1=J_+=\{i: x_i>0\}$ and
$S_2=J_{-}=\{i: x_i<0\}. $  Since $x$ can be any $K$-sparse vectors,
the above property holds for any disjoint subsets $S_1,
S_2\subseteq\{1,...,n\}$ with $|S_1|+|S_2| \leq K.$ Thus the RSP of
order $K$ holds. ~~ $\Box $

The theorem above shows that \emph{the RSP of order $K$ is a
necessary and sufficient condition for the exact recovery of all
$K$-sparse vectors.} Thus the RSP of order $K$ completely
characterizes the uniform recovery by  $\ell_1$-minimization.    It
is worth mentioning that Donoho \cite{D05} has characterized the
exact recovery condition from a geometric perspective, i.e., by the
so-called `$k$-neighborly' property.  Zhang \cite{Z08} has
characterized the uniform recovery by using null space property of
$A,$ and Juditsky and Nemirovski \cite{JN11}  have also proposed a
  condition for the uniform recovery based on
their function $\gamma (A). $ The RSP of order $K$ is an alternative
characterization of the uniform recovery.
 It is worth stressing that
although the RSP (\ref{range}) at an individual sparsest solution
does not imply the uniqueness of the sparsest solution (as shown in
section 3),    the RSP of order $K$  is more restrictive than the
individual RSP (\ref{range}). In fact, as indicated by the next
corollary, the RSP of order $K$ complies with the recovery
conditions (C1)--(C3).

\textbf{Corollary 4.3.} \emph{Let $A\in R^{m\times n} $ with $m<n.$
If $A^T$ has the RSP of order $K$, then any $\widetilde{x}\in R^{n}$
with $\|\widetilde{x}\|_0 \leq K $ is both the unique least
$\ell_1$-norm solution and  the unique sparsest solution to the
system $Ax=y=A\widetilde{x}.$}

\emph{Proof.} By Theorem 4.2, any $\widetilde{x}$ with
$\|\widetilde{x}\|_0\leq K$ can be exactly recovered by  $
\ell_1$-minimization. Thus, $\widetilde{x} $ is the unique least
$\ell_1$-norm solution. We now assume that there exists another
solution $x'$ to $Ax=A\widetilde{x}$  with $\|x'\|_0\leq
\|\widetilde{x}\|_0\leq K.$ Let $S_1=J'_+=\{i: x'_i>0\}$ and
$S_2=J'_{-}=\{i: x'_i<0\}.$ By the definition of the RSP of order
$K$, we see that the RSP (\ref{range}) holds at $x'$. Since any $K$
columns of $A$ are linearly independent (by Theorem 4.2), so are the
columns of $(A_{J'_+} ~ A_{J'_{-}}). $  Thus, by Theorem 2.10, $x'$
is the unique least $\ell_1$-norm solution. Therefore,
$x'=\widetilde{x},$ which shows that $\widetilde{x}$ is the unique
sparsest solution to the system $Ax=A\widetilde{x}.$ ~~ $\Box$

 Does a sensing matrix with the RSP
of certain order exist? The answer is obvious,
 as demonstrated by the next result, to which the proof is given in
 Appendix.

 \textbf{Lemma 4.4.} \emph{Suppose that  one of the following holds: }

 (i)  \emph{$K \leq \left\lfloor \frac{1}{2} \left(1+\frac{1}{\mu(A)}\right)\right\rfloor.$}

 (ii)   \emph{The matrix $A$ has the RIP of order $2K$ with constant $\delta_{2K}
 <\sqrt{2}-1.$}

 (iii)  \emph{The matrix $A$ has the NSP of order $2K.$}

\emph{Then the matrix $A^T$ has  the RSP of order $K.$}

 Therefore,  the existence of RSP matrices
is not an issue. In fact, any sufficient conditions for the uniform
recovery (not just those listed in Lemma 4.4) imply the RSP of
certain order.

\subsection{Beyond the uniform recovery}

Theorem 4.2  implies that it is impossible to uniformly recover all
$k$-sparse vectors with $k\geq \textrm{spark}(A)$ by a single matrix
$A.$ In order to recover a $k$-sparse vector with high sparsity, for
instance, $m> k>\frac{1}{2} \textrm{spark}(A), $  we should relax
the recovery condition (C2) which, according to Theorem 2.10, is not
a necessary condition for a vector to be exactly recovered by
$\ell_1$-minimization. Let us first relax this condition by dropping
the uniqueness requirement of the sparsest solution to the equation
$Ax=y,$ where $y$ denotes the measurements. Then we have the
following immediate result.

 \textbf{Proposition 4.5.} \emph{Let  $A$ satisfy the
following property: For any $k$-sparse vector $x^* $ with $Ax^*\not
=0,$ $x^*$ is a sparsest solution to the system $Ax=Ax^*.$ Then $ k
<spark(A).$}

\emph{Proof.} Note that for any sparsest solution $x$ to the  system
$Ax=b=Ax^*\not=0$, the columns
 of $A_{Supp(x)} $ are linearly independent (since otherwise, a sparser solution than $x$ can be found).
 Thus under the condition, we conclude that any $k$ columns of $A$
 are linearly independent.  So $k<\textrm{spark}(A).$   ~~ $\Box $

The proposition above shows that even if we relax  condition (C2) by
only requiring that $x^*$ be a sparsest
 solution (not necessarily the unique sparsest solution) to the system $Ax=y=Ax^*, $
  $\textrm{spark} (A)$ is
an unattainable upper bound for the uniform recovery.  This fact was
initially observed by Donoho and Elad \cite{DE03}, who had developed
the mutual coherence condition to guarantee the recovery success by
$\ell_1$-minimization.  Note that the uniform recovery by a matrix
with RIP or NSP of order $2k$ can recover all $k$-sparse vectors
with $k< \textrm{spark} (A)/2. $ Thus, from a mathematical point of
view, it is interesting to study how a
   sparse vector with $\|x\|_0\geq   \textrm{spark}(A)/2$ can be possibly recovered.  This
 is also motivated by some practical applications, where an unknown
 signal  might  not  be sparse enough to
fall into the range $\|x\|_0< \textrm{spark}(A)/2.$ Hence the
uniform recovery conditions (such as the RIP or NSP of order $2k$) does
not apply to these situations. Theorem 2.10  makes it possible to
handle such a situation by abandoning the recovering principle (C2).
This theorem shows that any solution, satisfying the individual RSP
(\ref{range}) and full-rank property, has a guaranteed recovery by
$\ell_1$-minimization. To satisfy these conditions, the targeted
signal does not have to be the sparsest solution to a linear system,
as shown by the next example.

\textbf{Example 4.6}. Let  $$A = \left(
  \begin{array}{rrrrr}
    6 & -4 & 3 & 4 & -2 \\
    6 & -4 & -1 & 4 & 0 \\
    0 & 2 & 3 & -1 & -3 \\
  \end{array}
\right), ~   y = \left(
                   \begin{array}{r}
                     4 \\
                     4 \\
                     -1 \\
                   \end{array}
                 \right).
$$
It is easy to check that $x^*= (1/3, -1/2, 0, 0, 0)^T $ satisfies
the RSP (\ref{range}) and full-rank property. Thus, by Theorem 2.10,
$x^*$ is  the unique least $\ell_1$-norm solution to the  system
$Ax=y.$  Thus $x^*$ can be exactly recovered by
$\ell_1$-minimization although it is not the sparsest one. It is
evident that $\widetilde{x}=(0, 0, 0, 1, 0)^T$ is the unique
sparsest solution (with $\|\widetilde{x}\|_0=1$) for this linear
system. (It is worth noting that $\widetilde{x}$
 cannot be
recovered since the RSP (\ref{range}) does not hold at this
solution.)

 Therefore Theorem 2.10 makes it possible to develop an
 extended uniform recovery theory. Toward this goal, we introduce the following
matrix property.

\textbf{Definition 4.7} (Weak-RSP of order $K$). \emph{Let $A\in
R^{m\times n}$  with $m<n.$   $A^T$ is said to satisfy the weak
range space property
  of order $K$ if (i) there exists a pair of disjoint subsets $S_1, S_2\subseteq \{1,..., n\}$
such that  $ |S_1|+|S_2|=
 K $ and  $\left(\begin{array}{cc}
      A_{S_1} & A_{S_2}
     \end{array}
   \right)$ has full column rank, and
(ii) for any disjoint $S_1, S_2\subseteq \{1,..., n\}$ such that $
|S_1|+|S_2|\leq
 K $ and  $\left(\begin{array}{cc}
      A_{S_1} & A_{S_2}
     \end{array}
   \right)$ has full column rank,  the space $ {\mathcal R}(A^T) $ contains a vector $\eta$ such
 that $\eta_i =1 $ for  $ i \in S_1,$  $ \eta_i =-1
 $ for  $ i\in
 S_2, $ and $ |\eta_i| <1 $ otherwise.}

The essential difference between this concept and the RSP of order
$K$ is that the RSP of order $K$ requires that the individual RSP
  hold for any disjoint subsets $S_1, S_2$ of $\{1, ..., n\} $ with $ |S_1|+|S_2|\leq K,$ but the Weak-RSP of order $K$
requires that the individual RSP   hold only  for those disjoint
subsets $S_1, S_2$ satisfying that $|S_1|+|S_2|\leq K$ and
$\left(\begin{array}{cc}
      A_{S_1} & A_{S_2}
     \end{array}
   \right)$ has full-column-rank.  So the RSP of order $K $ implies  the Weak-RSP of $K,$
   but the converse is not true in general.   Based on this concept,
   we have the next  result that follows
 from Theorem 2.10 immediately.

 \textbf{Theorem 4.8. } (i) \emph{If
$A^T$ has the Weak-RSP of order $K, $ then $K\leq  m.$}

(ii) \emph{Assume that the measurements of the form $y= Ax$ are
taken. Then any $x,$ with $\|x\|_0 \leq K$ and
$\left(\begin{array}{cc}
      A_{J_+} & A_{J_{-}}
     \end{array}
   \right)$ being full-column-rank, can be exactly
recovered by $\ell_1$-minimization if and only if $A^T$ has the
Weak-RSP of order $K.$ }

The bound $K\leq m$ above follows directly from the condition (i) of
Definition 4.7. It is not difficult to see a  remarkable difference
between Theorems 4.8 and 4.2.  Theorem 4.2 claims that all vectors
with sparsity $\|x\|_0\leq K$ can be exactly recovered via a sensing
matrix with the RSP of order $K$, where $K< \textrm{spark}(A)$ which
is an unattainable upper bound for any uniform recovery. Different
from this result, Theorem 4.8 characterizes the exact recovery of a
part (not all) of  vectors within the  range $1\leq \|x\|_0\leq m.$
This result makes it possible to use a matrix with the Weak-RSP of
order $K,$ where $ \textrm{spark}(A)/2 \leq K < m,$ to exactly
recover some sparse vectors in the range $ \textrm{spark}(A)/2 \leq
\|x\|_0 < K,$ to which the current uniform-recovery theory is
difficult to apply.

It is worth stressing that the Weak-RSP-based recovery has
abandoned the
   uniqueness requirement (C2) of the sparsest solution to a  linear
   system, and has built the recovery theory
    on  condition (C1) only. Thus the guaranteed recovery can be naturally
   extended into the range
    $[\textrm{spark}(A)/2, m).$ Of course, only
some  vectors (signals) in this range can be exactly recovered,
i.e., those vectors with the RSP (\ref{range}) and the full-rank
property. Both Theorems 4.2 and 4.8  have shed light on the limit of
the recovering ability of $\ell_1$-minimization.

Finally, we point out that although checking the RSP at a given point $x$ is easy,  checking
the RIP, NSP,  and RSP of certain order for a matrix is generally
difficult.  From a practical point of view, it is also important to
develop some verifiable conditions (see e.g., \cite{JN11, TN11}).

\section{Conclusions}
The uniqueness of  least $\ell_1$-norm solutions to  underdetermined
linear systems plays a key role in solving $\ell_0$-minimization
problems and in sparse signal recovery. Combined with Fuchs'
theorem,  we have proved that a vector is the unique least
$\ell_1$-norm solution to a linear system if and only if the
so-called range space property and full-rank property hold at this
vector. This complete characterization provides immediate answers to
several questions in this field. The main results in this paper were
summarized in Theorems 2.10, 3.3, 4.2 and 4.8. These results have
been developed naturally from the classic linear programming theory,
and have been benefited by distinguishing between the equivalence
and strong equivalence of $\ell_0$- and $\ell_1$-minimization. The
RSP-based analysis in this paper is useful to explore a broad
equivalence between $\ell_0$- and $\ell_1$-minimization, and to
further understand the internal mechanism and capability of the
$\ell_1$-method for solving $\ell_0$-minimization problems.
Moreover, we have introduced such new matrix properties as the RSP
of order $K$ and the Weak-RSP of order $K. $ The former turns out to
be one of the mildest conditions governing the uniform recovery,
 and the latter may yield an
extended uniform recovery.

It is worth mentioning that the discussion in this paper was focused
on the sparse signal recovery without noises.  Some open questions
are worthwhile to address in the future, such as how the RSP can be
used to analyze the sparse signal recovery with noises,  and how the
RSP-based analysis can be possibly used to establish a lower bound
for measurements in compressed sensing.

\vskip 0.2in \noindent Appendix \\

 \emph{Proof of Lemma 2.2:} The equivalence between (i) and (ii) is obvious. The equivalence
between (ii) and (iii) can be easily verified as well. Suppose that
(ii) holds.  We now show that $(u,t)= (0, |x|)$ is the only solution
to (\ref{system}). In fact, let $(u,t)$ be an arbitrary solution to
(\ref{system}).  Since $ |u_i+x_i| \leq t_i$ for all  $ i=1, ...,
n,$ adding all these inequalities leads to $\|u+x\|_1 \leq
\sum_{i=1}^n t_i$ which, combined with the second inequality of
(\ref{system}), yields $\|u+x\|_1\leq \|x\|_1.$ This implies that
$u\in C\bigcap {\cal N}(A).  $ Since $C\bigcap {\cal N}(A)=\{0\},$
we must have $u=0.$ Substituting $u=0$ into (\ref{system}), we see
that $t$ must be $|x|.$  In other words, under (ii), $(u,t)= (0,
|x|) $ is the only solution to (\ref{system}).  The converse can
be  verified easily as well.\\

\emph{Proof of Lemma 2.3:}     Note that slack variables of ($LP_2$)
are uniquely determined
   by $(u,t)$ as follows:  \begin{equation} \label{111} \alpha= t-x-u  , ~~
   \beta  = t+x+u, ~~ r=\|x\|_1-\sum_{i=1}^n t_i. \end{equation}
There is a one-to-one correspondence between  feasible points of
($LP_1$) and ($LP_2$), i.e., $(u,t)$ is feasible to ($LP_1$) if and
only if $ (u, t, \alpha, \beta, r),$ where $(\alpha, \beta, r)$ is
given by (\ref{111}),  is feasible to ($LP_2$).  Since both problems
have zero objectives, any feasible point is optimal. Thus (i) and
(ii) are equivalent. Also, there exists
 a one-to-one correspondence between  feasible points of ($LP_2$) and
($LP_3$). In fact, it is evident that $(u, t, \alpha,\beta, r)$ is
feasible to ($LP_2$) if and only if $(u', t, \alpha, \beta, r)$ is
feasible to ($LP_3$), where $u'=Me-u \geq 0$ (the nonnegativity
follows from the definition of $M$).  Note  that both problems have
zero objectives. Thus (ii) and (iii) are also equivalent. \\

 \emph{Proof of Lemma 2.5: } First, we assume that   $(y,y',  \omega)$
satisfies (\ref{condition}). Set $\eta= (y-y')/\omega.$ We
immediately see that $\eta\in {\mathcal R}(A^T),$ and  $\eta_i=
(y_i-y'_i)/\omega =1  $ for every $x_i > 0$ (since $y_i=\omega$ and
$y_i'=0$ for this case).    Similarly we have $\eta_i=-1$ for every
$i$ with $x_i<0.$ For $x_i=0$, since
 $\omega < y_i+y_i'$ and both $y_i$ and $y'_i$ are negative, it follows
 that
 $|\eta_i| =  |y_i-y_i'|/|\omega| < |y_i+y_i'|/ |\omega| <1.$

 Conversely, assume that there is a vector $\eta \in {\mathcal R}(A^T) $
such that $
 \eta_i = 1 $   for  all $ x_i>0,$
   $\eta_i =-1 $ for all $  x_i<0 , $ and $
  |\eta_i| < 1$ for  all $  x_i =0.$   We now construct a vector
 $(y,y',  \omega) $ satisfying  (\ref{condition}). Indeed, let us first set $\omega=-1,$ and then set
 $y_i=0,  ~y'_i=-1  $ for $ x_i<0, $ and
 $y_i=-1, y_i'=0 $ for $ x_i>0. $
 For those $i$ with $x_i=0$, since $|\eta_i|<1,$   there exists a
 constant $\varepsilon_i$ such that
   $0< \varepsilon_i
 <(1- |\eta_i|)/2,$  and thus we define   $y_i $ and $ y'_i$ as
 follows:
 \begin{equation} \label{yyy} \left\{ \begin{array} {ll} y_i =
 -\varepsilon_i-\eta_i \textrm{ and } y'_i =  -\varepsilon_i
  & \textrm{ if }
 \eta_i>0, \\
 y_i = -\varepsilon_i \textrm{ and } y'_i = \eta_i-\varepsilon_i   &
\textrm{ otherwise.}\end{array} \right.
 \end{equation}
From the above construction,   it is easy to see that $y-y'=-\eta.$
Thus $y-y'\in {\mathcal R}(A^T).$  To verify that $(y, y', \omega) $
satisfies  (\ref{condition}), it is sufficient to show that $
-1=\omega < y_i+ y'_i, ~y_i<0, ~y'_i<0 $ for all $ x_i=0.$ Indeed,
we see from (\ref{yyy}) that both $y_i $ and $ y'_i$ are negative,
and
$$ |y_i+y'_i|=\left\{\begin{array} {ll} |(-\varepsilon_i-\eta_i) +
(-\varepsilon) | \leq 2\varepsilon_i+  |\eta_i|  & \textrm{ if }\eta_i>0\\
|(-\varepsilon)+  (\eta_i-\varepsilon_i )| \leq 2\varepsilon_i+
 |\eta_i|   & \textrm{ otherwise, } \end{array}\right. $$
 which by the definition of $\varepsilon_i$ implies that
 $|y_i+y'_i|<1.$ Since $y_i<0$ and $y'_i<0,$ this implies that  $0> y_i +y'_i > -1 =
\omega.$ Thus $(y, y', \omega)$ constructed as above does satisfy
  (\ref{condition}). \\

\emph{Proof of Lemma 4.4:} The mutual-coherence condition implies
that any $x$ with
 $ \|x\|_0 <   (1+1/\mu(A))/2 $   is
 both the unique sparsest and the unique
 least $\ell_1$-norm solutions to
  the  system $Az=y=Ax. $ Let $S_1= \{i: x_i>0\} $ and $ S_2=
  \{i: x_i<0\}. $ By Theorem
  2.8, there exists a vector $\eta \in {\cal R} (A^T) $ satisfying the RSP (\ref{range}) at $x,$ i.e.,
    $\eta_i=1 $ for $i\in S_1,
  $ $\eta_i=-1$ for $i\in S_2,$ and $|\eta_i|<  1 $ otherwise.  Since
  $x$ here can be any sparse vector with $ \|x\|_0 \leq
  K^0=: \left\lfloor \frac{1}{2} \left(1+\frac{1}{\mu(A)}\right)\right\rfloor,
  $ the above-defined $S_1 $ and $S_2$ can be  any disjoint subsets $S_1$, $S_2$ of $\{1, ..., n\}$ with
  $|S_1|+|S_2| \leq K^0.$   Thus  $A^T$ has the RSP of at least order $K^0.$
 Both the RIP  and NSP of order $2K$ imply that any sparse vector
$x$ with $\|x\|_0\leq K$ is the unique sparsest solution and the
unique least $\ell_1$-norm solution to the system $Az=y =Ax.$  By
the same analysis  above, it implies that the RSP of order $K$
holds.

\begin{IEEEbiographynophoto} {Yun-Bin Zhao} Yun-Bin Zhao received his PhD degree from the
Institute of Applied Mathematics, Chinese Academy of Sciences (CAS)
in 1998. From June 1998 to February 2001 he was a postdoctoral
research fellow with the Institute of Computational Mathematics,
CAS, and with the Department of Systems Engineering \& Engineering
Management, Chinese University of Hong Kong. From 2001 to 2003, He
was an assistant professor in the Academy of Mathematics and Systems
Science (AMSS), CAS, and from 2003 to 2007 he was an associate
professor in AMSS. He joined the University of Birmingham  in
2007,  as a lecturer of mathematical optimization. Since 2013, he is
a senior lecturer in the same university. He serves as an associate
editor of \emph{Applied Mathematics and Computation}. His research
interests include the operations research, computational
optimization, convex analysis, numerical linear algebra, and their
applications.
 \end{IEEEbiographynophoto}

 \end{document}